\documentclass[copyright,creativecommons]{eptcs}

\usepackage{graphicx}

\usepackage{pdfpages}

\usepackage{float}

\usepackage{amssymb}
\usepackage{amstext}
\usepackage{amsmath}

\providecommand{\event}{ThEdu'18 Post-Proceedings} % Name of the event you are submitting to
\usepackage{underscore}           % Only needed if you use pdflatex.

\title{Students' Proof Assistant (SPA)}
\author{Anders Schlichtkrull \email{} \and J{\o}rgen Villadsen \email{} \and Andreas Halkj{\ae}r From \email{}
\institute{DTU Compute - Department of Applied Mathematics and Computer Science,\\[1ex]
Technical University of Denmark, Richard Petersens Plads, Building 324, DK-2800 Kongens Lyngby, Denmark}
}
\def\titlerunning{Students' Proof Assistant (SPA)}
\def\authorrunning{A. Schlichtkrull, J. Villadsen \&\ A. H. From}

\begin{document}

\maketitle

\begin{abstract}
\noindent
The Students' Proof Assistant (SPA) aims to both teach how to use a proof assistant like Isabelle and also to teach how reliable proof assistants are built. Technically it is a miniature proof assistant inside the Isabelle proof assistant. In addition we conjecture that a good way to teach structured proving is with a concrete prover where the connection between semantics, proof system, and prover is clear. The proofs in Lamport's TLAPS proof assistant have a very similar structure to those in the declarative prover SPA. To illustrate this we compare a proof of Pelletier's problem 43 in TLAPS, Isabelle/Isar and SPA. We also consider Pelletier's problem 34, also known as Andrews's Challenge, where students are encouraged to develop their own justification function and thus obtain a lot of insight into the proof assistant.
Although SPA is fully functional we have so far only used it in a few educational scenarios.
\end{abstract}

\thispagestyle{empty}

\section{Introduction}

Our Students' Proof Assistant (SPA) aims to both teach how to use a proof assistant like Isabelle \cite{nipkow+02} and to teach how reliable proof assistants are built. SPA is a miniature proof assistant running inside Isabelle (\url{https://github.com/logic-tools/spa}). It is based on work by Harrison \cite{harrison+09} and the entire development runs in Isabelle's ML environment as an interactive application. The details are in our recent publication \cite{jensen2018}. In that publication we formalized the kernel of an axiomatic system in Isabelle and exported it to SML-code. This code served as the kernel of a translation from OCaml to SML of a proof assistant from Harrison's chapter on the topic \cite{harrison+09}.

\begin{figure}
Incomplete Proof:
\smallskip
\begin{center}
\includegraphics[width=\textwidth,height=\textheight,keepaspectratio]{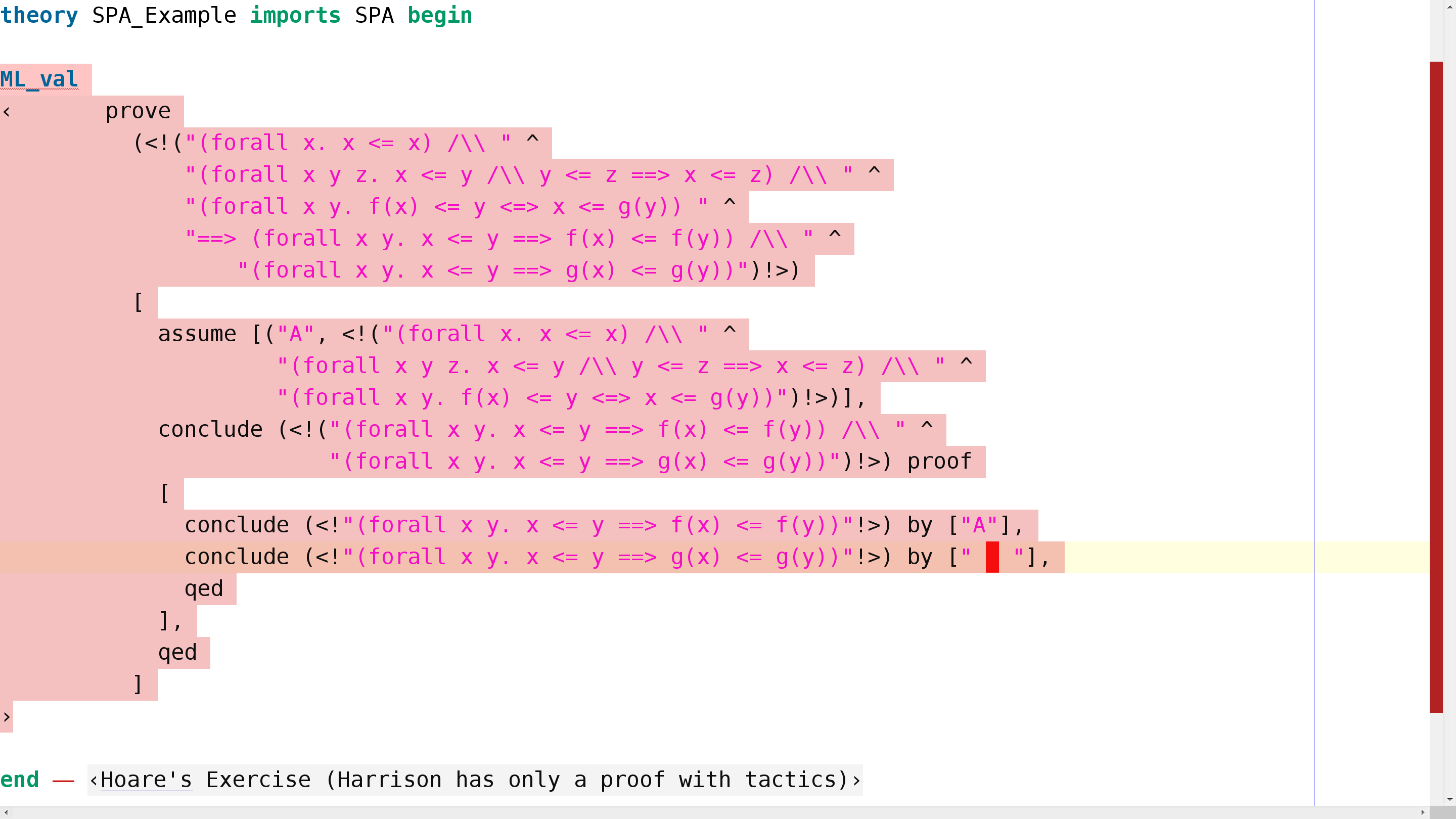}
\end{center}
\bigskip\par\bigskip
Complete Proof:
\smallskip
\begin{center}
\includegraphics[width=\textwidth,height=\textheight,keepaspectratio]{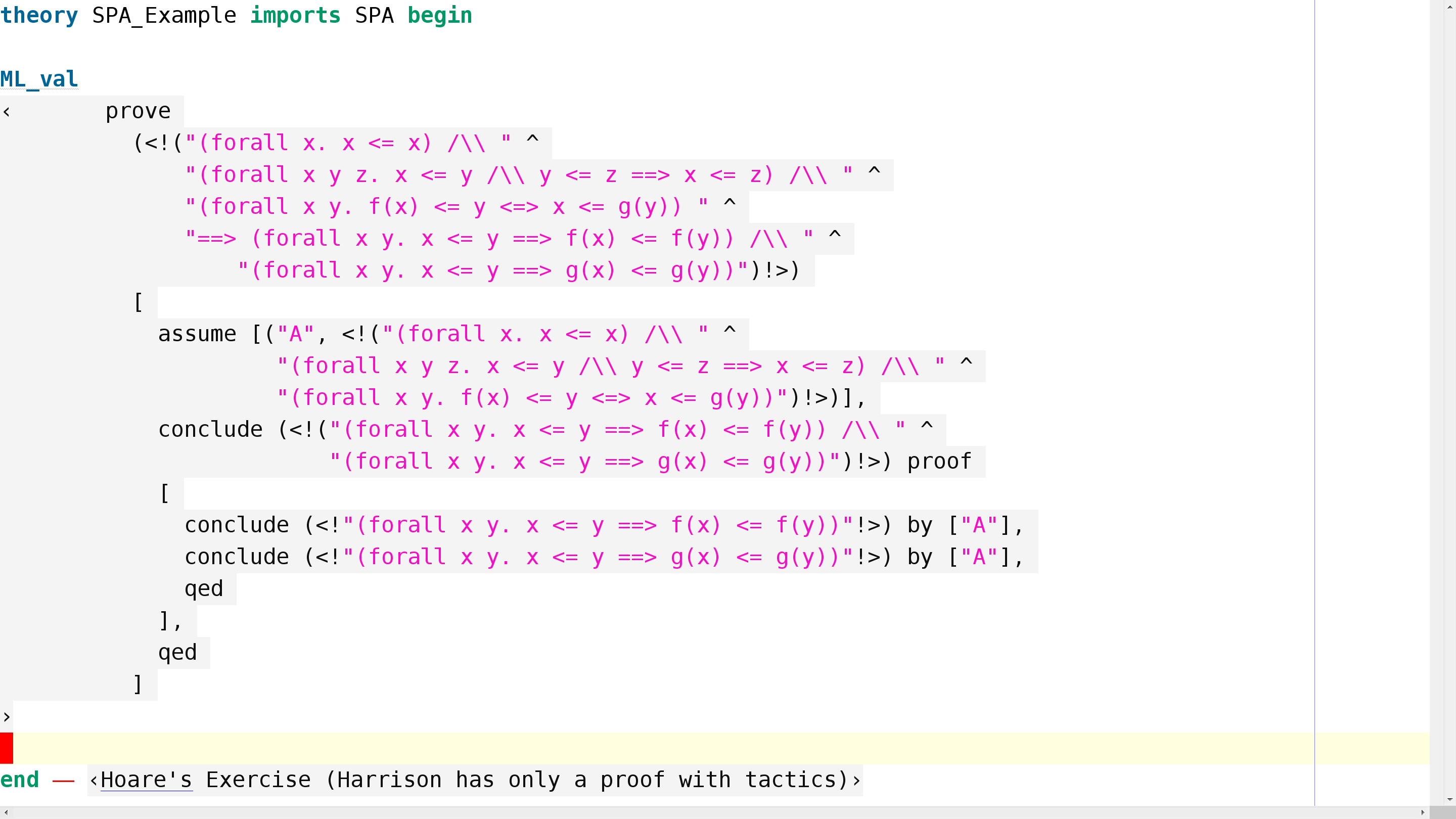}
\end{center}
\caption{Proof Example in Students' Proof Assistant (SPA) (Isabelle Screenshots) \label{ex}}
\end{figure}

In the present paper we first describe the main changes we made to Harrison's code since our publication \cite{jensen2018} --- resulting in SPA --- and then we consider the proofs of Pelletier's problems 43 and 34 that we use in our logic teaching. Figure \ref{ex} shows a proof example (Isabelle screenshots of an incomplete proof as well as a complete proof) of the formula in ``On an exercise of Tony Hoare's'' by Edsger W. Dijkstra \cite{EWD:EWD1062}.
In the incomplete version, the reference to the assumption ``A'' is missing in the justification, so the conclusion cannot be proven.
Due to the current integration between SPA and Isabelle, this error in a part of the proof causes the entire proof to become highlighted as incomplete.
This showcases how it currently feels to work with the integrated proof assistant, an experience we wish to improve in the future.
This way of specifying the use of assumptions mimics many proof assistants, e.g. Isabelle, so students familiar with it can apply it elsewhere.

We conjecture that a good way to teach structured proving is with a concrete prover like SPA where the connection between semantics, proof system, and prover is clear. Even for paper proofs Lamport recommends writing in a structured style \cite{lamport,lamport2}.
Although SPA is fully functional we have so far only used it in a few educational scenarios but we are planning a new course on automated reasoning where SPA will play a central role.

\newpage

\section{Main Changes to Harrison's Code}

In our publication \cite{jensen2018} we chose to let the definitions in the formalization and translation follow Harrison's book \cite{harrison+09} very strictly. Harrison's code works well in a book on the broad topic of practical logic and automated reasoning but is in places perhaps overly general for teaching material only on the topic of proof assistants. For instance Harrison's datatype for formulas is parameterized with the type for atoms: for first-order logic the parameter is instantiated to a type for first-order predicates, and for propositional logic the parameter is instantiated with the strings and ignores the constructors for universal and existential quantification. 
Since we will only consider a proof assistant for first-order logic we do not need to parameterize on a type of atoms -- we can put first-order predicates directly into the definition. We see more opportunities for improving the development for our purpose.

Our previous publication \cite{jensen2018} discussed three ways to represent the type of theorems:
\begin{enumerate}
\item A datatype wrapping the type of formulas in a constructor.
\item A type exactly characterizing the provable formulas (theorems).
\item A type exactly characterizing the valid formulas.
\end{enumerate}
In the first case the theorems were then further characterized by a predicate. The latter two cases can be made using Isabelle's \textbf{typedef} command as well as functionality for lifting functions that work on formulas to work on the new type.
We previously argued for 1 because of its simplicity, but for SPA we have had a change of hearts and instead go with solution 2. The reason is that while lifting and \textbf{typedef} are arguably advanced concepts they make it easier to inspect the verification as we argued \cite{jensen2018} and furthermore the idea behind \textbf{typedef} is after all quite simple -- it can be seen as elevating a set to a type.

\section{Pelletier's Problem 43}

\begin{figure}
\begin{center}
\includegraphics[trim=20mm 60mm 20mm 12mm,clip,width=\textwidth,height=\textheight,keepaspectratio]{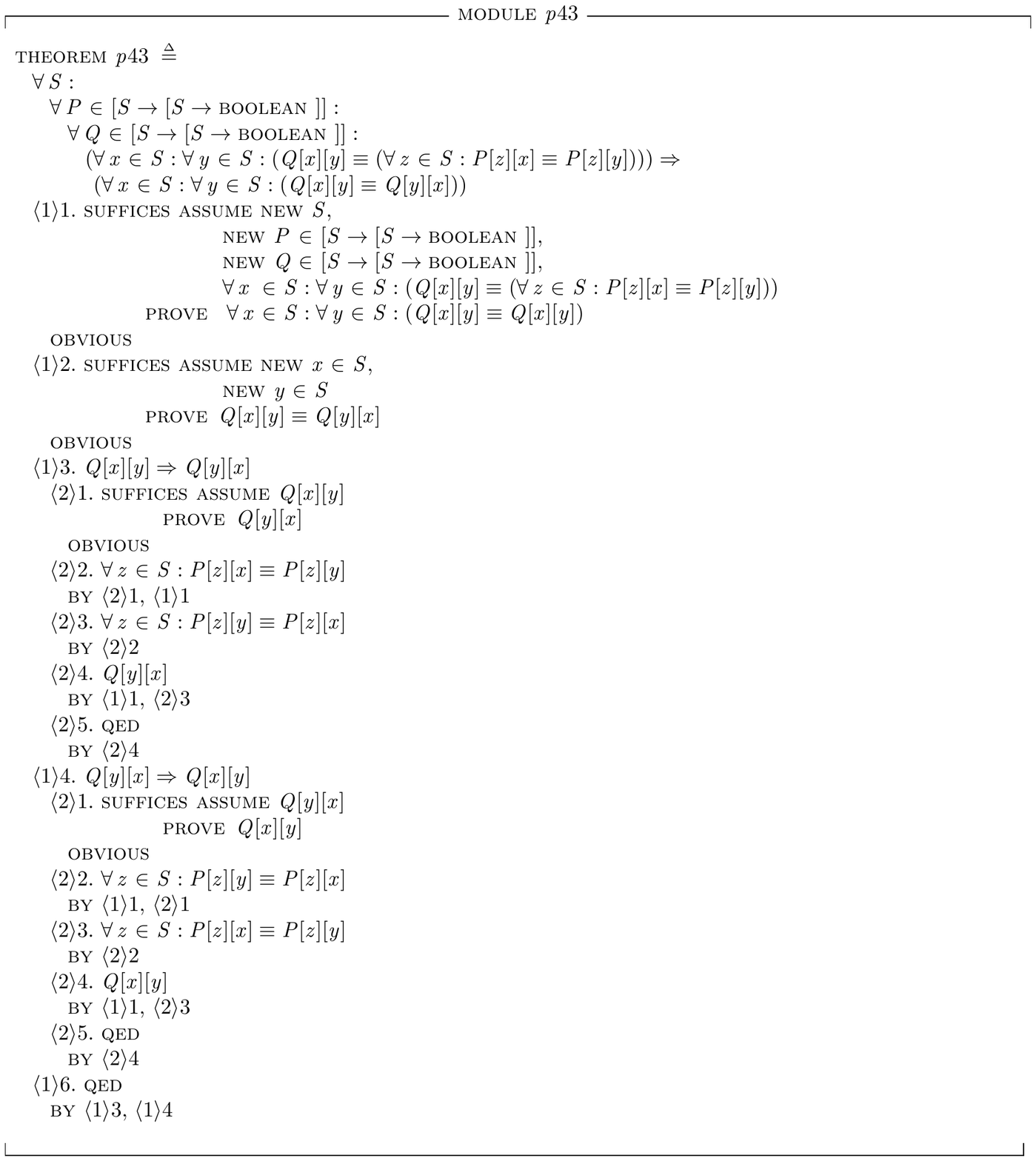}
\end{center}
\caption{Proof of Pelletier's Problem 43 in the TLA+ Proof System (TLAPS) \label{figuretlaps}}
\end{figure}

\begin{figure}
\begin{center}
\includegraphics[trim=10mm 70mm 10mm 10mm,clip,width=\textwidth,height=\textheight,keepaspectratio]{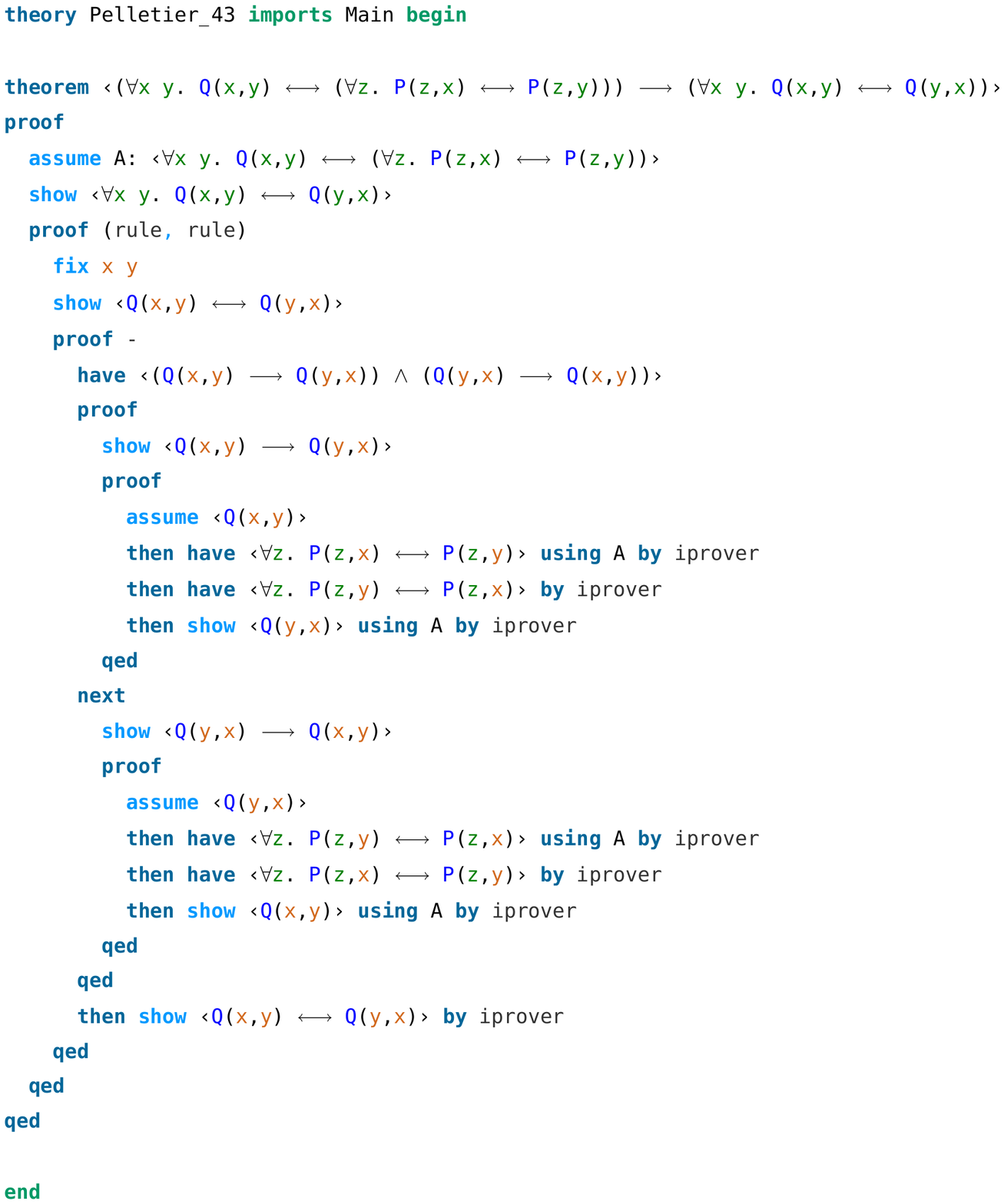}
\end{center}
\caption{Proof of Pelletier's Problem 43 in Isabelle/Isar\label{figureisabelleisar}}
\end{figure}

\begin{figure}
\begin{center}
\includegraphics[trim=3mm 42mm 3mm 15mm,clip,width=\textwidth,height=\textheight,keepaspectratio]{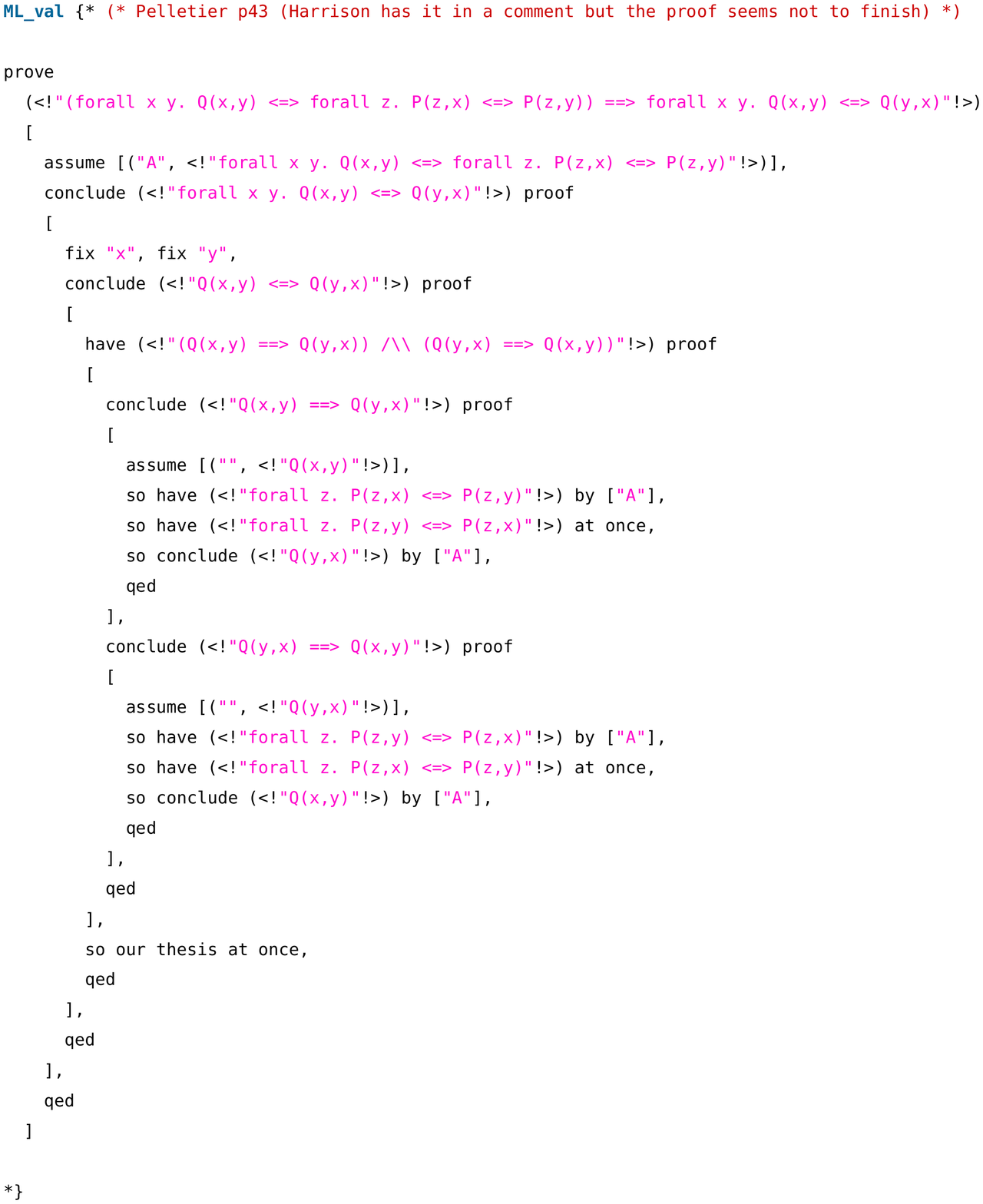}
\end{center}
\caption{Proof of Pelletier's Problem 43 in Students' Proof Assistant (SPA) \label{figurespa}}
\end{figure}

The proofs in Lamport's TLAPS proof assistant have a very similar structure to those in the declarative prover SPA.
To illustrate this we compare a proof of Pelletier's problem 43 \cite{jensen2018} in TLAPS (Figure \ref{figuretlaps}), Isabelle/Isar (Figure \ref{figureisabelleisar}) and SPA (Figure \ref{figurespa}). 

Pelletier's problem 43 is:
\[
(\forall  x \; y. \; Q(x,y) \longleftrightarrow (\forall z. \; P(z,x) \longleftrightarrow P(z,y)) )
\longrightarrow
(\forall  x \; y. \; Q(x,y) \longleftrightarrow Q(y,x))
\]
The idea is that based on a binary relation $P$ the relation $Q$ is defined to consist of any pair $(x,y)$ that has the property 
\[ 
\forall z. \; P(z,x) \longleftrightarrow P(z,y)
\]
In other words, any $x$ and $y$ are related by $Q$ if they relate equivalently to any $z$ with regards to $P$.
The problem states that $Q$ is symmetric.

We explain the SPA proof informally, using regular notation:

\medskip

\noindent
\textbf{Proof:}\\
We are trying to prove an implication, so we start off by assuming the antecedent, calling it ``A'' so we can refer to it later in the proof.
Since the statement is universally quantified, we arbitrarily fix an $x$ and a $y$ to use in the proof.
We then show the bi-implication by showing the conjunction of both directions and using the command \textit{at once} which does pure first-order reasoning and can easily handle small steps such as this one.

Consider first the direction $Q(x, y) \longrightarrow Q(y, x)$.
This is an implication so we start again by assuming the antecedent.
We do not have to name it this time, as we only use it in the proof of the next statement where it can be referenced using \textit{so}.
This assumption matches the left-hand side of ``A'' which we appeal to using \textit{by} and are thus allowed to conclude the right-hand side: That $x$ and $y$ are equivalent with regards to $P$.
The next line swaps the order of the bi-implication \textit{at once}, which allows us to then appeal to ``A'' again, this time in the opposite direction, and conclude the goal $Q(x, y)$.
Note that in the final step, the quantified $x$ in ``A'' is instantiated with our fixed $y$ and $y$ with $x$.

The proof of the direction $Q(y, x) \longrightarrow Q(x, y)$ is exactly symmetric to the one above.
\\\makebox[\textwidth][r]{$\square$}

\

We mark the end of a (sub)proof by \textit{qed} in TLAPS, Isabelle/Isar and SPA.

In TLAPS, the types of predicates must be explicitly given, while these are inferred in Isabelle/Isar and SPA.
We argue that this reduces the number of concepts students need to understand to get started.
Furthermore, the syntax used for predicate application in SPA matches closer the syntax usually used in textbooks.

In our opinion, the use of pure ASCII in SPA is an advantage in a setting where students are not used to programming with anything else.
The use of Greek letters or mathematical symbols directly in the proof code can then be added later, when convenient.

\section{Pelletier's Problem 34}

Amongst Pelletier's problems another interesting one is problem 34 which is also known as Andrews's Challenge \cite{harrison+09}. The truth of the formula is not obvious at first glance since it relies on the fact that bi-implication is both commutative and, perhaps surprisingly, associative. We have proved the formula in SPA and the about 150 lines are listed in the Appendix.

\smallskip
\begin{quote}
``At the Marktoberdorf summer school in August 1996, Larry Paulson lectured on his mechanical theorem prover, Isabelle; Natarajan Shankar lectured on his mechanical theorem prover, PVS; and I lectured on calculational logic. Both Paulson and Shankar suggested that I try the calculational approach on Andrew's challenge, which is one of several difficult theorems used to benchmark mechanical theorem provers.''

\smallskip
\hfill Davis Gries \url{https://www.cs.cornell.edu/gries/Papers/andrews.html}
\end{quote}
\smallskip

\noindent
We set out to prove Pelletier's problem 34:
\begin{align*}
  & ((\exists x. \forall y. P(x) \leftrightarrow P(y)) \leftrightarrow ((\exists x. Q(x)) \leftrightarrow (\forall y. Q(y))))
  \leftrightarrow \\
  & ((\exists x. \forall y. Q(x) \leftrightarrow Q(y)) \leftrightarrow ((\exists x. P(x)) \leftrightarrow (\forall y. P(y))))
\end{align*}
The automated prover is not powerful enough to prove this formula automatically so we need to manually break it down into smaller components until they are so small that the prover is powerful enough to take over and then reassemble these proofs into a proof of the original formula.
For instance, we prove the outer bi-implication by proving each direction separately and showing that these imply the bi-implication.

Two axioms in particular are useful for this proof, namely the one which allows us to read a bi-implication from left to right:
\[ \vdash (p \leftrightarrow q) \rightarrow (p \rightarrow q) \tag{\texttt{axiom\_iffimp1}} \]
And the one allowing us to form a bi-implication from two implications:
\[ \vdash (p \rightarrow q) \rightarrow (q \rightarrow p) \rightarrow (p \leftrightarrow q) \tag{\texttt{axiom\_impiff}}\]
Furthermore the derived rule \texttt{unshunt} is useful:
\[ \text{if } \vdash p \rightarrow q \rightarrow r \text{ then } \vdash p \land q \rightarrow r \tag{\texttt{unshunt}} \]
Finally we use the tactic \texttt{conj\_intro\_tac} to show a conjunction by showing its conjuncts.

Unfortunately, even when we break down the formula using these rules the automation gets stuck in a very basic step of the proof of
\[ (\exists x. \forall y. Q(x) \leftrightarrow Q(y)) \]
under the two assumptions:
\begin{equation}\label{assump1}
  (\exists x. P(x)) \leftrightarrow (\forall y. P(y)) \tag{*}
\end{equation}

\begin{equation}\label{assump2}
(\exists x. \forall y. P(x) \leftrightarrow P(y)) \leftrightarrow ((\exists x. Q(x)) \leftrightarrow (\forall y. Q(y))) \tag{**}
\end{equation}
From the first assumption~(\ref{assump1}) we can automatically derive
\[ \exists x. \forall y. P(x) \leftrightarrow P(y) \]
This matches the left-hand side of our second assumption~(\ref{assump2})
and we therefore expect to be able to conclude its right hand side.
We start by rewriting~(\ref{assump2}) into an implication from left to right. To do this we want to use an instance of \texttt{axiom\_iffimp1}:
\begin{align*}
  & (\exists x. \forall y. P(x) \leftrightarrow P(y)) \leftrightarrow ((\exists x. Q(x)) \leftrightarrow (\forall y. Q(y))) \rightarrow \\
  & (\exists x. \forall y. P(x) \leftrightarrow P(y)) \rightarrow ((\exists x. Q(x)) \leftrightarrow (\forall y. Q(y)))
\end{align*}
And now we would hope to conclude the implication on the bottom line as it follows directly from modus ponens and~\ref{assump2}.
Alas, the automatic prover takes a wrong turn and never seems to finish.
To mitigate this we write our own function to eliminate the implication manually.
We will reserve the word tactic to refer to functions that transform the current goals.
This function, however, produces a theorem instead of new goals, as such we call it a justification (function).

We write the justification as an ML function \texttt{by\_mp}.
Before showing its code we will give the intuition of how it works.
Following previous work~\cite{jensen2018}, we will depict justifications as schemas on the following form:
\smallskip
\[
    \left(
    \begin{matrix}
        \vdash p_{11} \land \ldots \land p_{1i_1} \rightarrow & q_1 \\
        \vdots & \\
        \vdash p_{n1} \land \ldots \land p_{ni_n} \rightarrow & q_n \\
    \end{matrix}
    \right)
    \Longrightarrow\ \vdash P
\smallskip
\]
This represents a justification function that derives the goal (here \(P\)) given proofs of the subgoals under their respective assumptions (depicted between the parentheses).

The justification function \texttt{by\_mp} is depicted as follows:
\smallskip
%\scalebox{0.7}{
\[
\left(
\begin{array}{lll} 
\vdash a \land (a \longrightarrow b) \land p_{01} \land \ldots \land p_{0i_0} &\longrightarrow & q_0 \\ 
\vdash p_{11} \land \ldots \land p_{1i_1} & \longrightarrow & q_1 \\ 
\vdots &  & \\
\vdash p_{n1} \land \ldots \land p_{ni_n} & \longrightarrow & q_n
\end{array} 
\right) 
\Longrightarrow \; \vdash a \land (a \longrightarrow b) \land p_{01} \land \ldots \land p_{1i_0} \longrightarrow b
\smallskip
\]
%}
In short, the idea is that from a state where we have available assumptions $a$ and $a \rightarrow b$ we can get to $b$ while keeping all the available assumptions.

Let us now explain \texttt{by\_mp} in detail. The function takes three arguments and produces the theorem we want.
The first argument is a pair of labels denoting respectively the implication and the antecedent that we want to use in our application of modus ponens.
The second argument is the current goal and the last argument stores a data type \texttt{goals} consisting of two parts: a list of subgoals and a justification function.
The ML-code of \texttt{by\_mp} is

\begin{verbatim}
fun by_mp (ab, a) p (Goals ((asl,_)::_, _)) =
  let
    val ths = assumps asl
    val th = right_mp (assoc ab ths) (assoc a ths)
    handle Fail _ =>
      raise Fail "by_mp: unapplicable assumptions"
  in
    if consequent (concl th) = p
    then [th]
    else raise Fail "by_mp: wrong conclusion"
  end
\end{verbatim}

\noindent
The function starts by looking up the implication and antecedent among the available assumptions (this is a slight generalization from the above depiction where they were assumed to be the first two assumptions available).
To ensure correctness, the \texttt{assumps} function turns each individual assumption into an implication from the conjunction of all of the assumptions to that assumption.
So in the context of conjoined assumptions \(p\) the implication \(q \rightarrow r\) becomes \(p \rightarrow q \rightarrow r\).
Similarly, our antecedent \(q\) is only available as \(p\rightarrow q\).
We use the transitivity rule \texttt{right\_mp} to apply modus ponens under these ambient assumptions:
\[
  \text{if } \vdash p \rightarrow q \rightarrow r
  \text{ and } \vdash p \rightarrow q
  \text{ then } \vdash p \rightarrow r
  \tag{\texttt{right\_mp}}
\]
Then, if the consequent \(r\) matches our input goal \(p\) the theorem we have produced is an implication from the ambient assumptions to the goal and can therefore be directly applied to solve that goal.
Otherwise we present the user with an error, as we do if the assumptions do not align.

Using \texttt{by\_mp} we can now get on with the proof by successfully concluding
\[ (\exists x. \forall y. P(x) \leftrightarrow P(y)) \rightarrow ((\exists x. Q(x)) \leftrightarrow (\forall y. Q(y))) \]
And since the left-hand side matches what we derived earlier we can finish the subproof by concluding the wanted
\[ ((\exists x. Q(x)) \leftrightarrow (\forall y. Q(y))) \]
The rest of the proof proceeds in a similar fashion and has three other cases where \texttt{by\_mp} is needed.

At first glance it seems annoying that the automatic prover cannot solve the presented case, and investigating whether this can be remedied might be fruitful, but writing our own justification function has in itself provided us with a lot of insight into the proof assistant.

First off, the actual programming and experimenting required is facilitated by the proof assistant's embedding in Isabelle: The ML code is automatically compiled and made available to the subsequent proofs.

This embedding also makes it easier to investigate how the existing proof-manipulating functions are constructed and how to write the new one.
All the ML code is available in a single file that you can navigate and errors are provided directly in the editor.

Finally the implementation of the function itself is like a little logic puzzle that invites the user to investigate what pieces are available --- axioms and derived rules --- and how to put them together.

As such, asking students to write not only their own proofs but also justifications or tactics might teach them a whole lot about the workings of the proof assistant through exploration.

\section{Discussion}

The similarities between the proofs in TLAPS, Isabelle/Isar and SPA are evident. This corroborates that the skills that the students learn in our miniature proof assistant are transferable to full-fledged proof assistants.

For Pelletier's problem 43 the automation in a proof assistant like Isabelle can actually prove the whole formula without the user supplying a proof. This raises the question: Why bother teaching students how to prove such lemmas in more detail than a single call to the automation? Our answer is that when working in a proof assistant for higher-order logic one often works by breaking down the logical structure of the problem. This often amounts to first-order reasoning. Hereafter one takes a look at the involved mathematical objects and tries to come up with a needed lemma involving these objects. Doing this of course requires knowledge of breaking the logical structure down to begin with. 

We have not found related work where a miniature proof assistant is developed inside another proof assistant. 
Overall our approach is related to the IsaFoL --- Isabelle Formalization of Logic (\url{https://bitbucket.org/isafol}) project which unites researchers in formalizing logic in Isabelle \cite{IsaFoL}. Among the formalizations in the project are SAT-solving, first-order resolution, a paraconsistent logic, natural deduction, sequent calculi and more.

\section{Conclusion}

SPA is an advanced e-learning tool for teaching proof assistants for students in computer science as well as in mathematics and complements our other e-learning tool, NaDeA (A Natural Deduction Assistant with a Formalization in Isabelle), which is available online and has been used by computer science bachelor students in regular courses at DTU \cite{DBLP:journals/corr/abs-1803-01473}. 

We have a number of ideas for improving SPA such as tighter integration between the miniature proof assistant and the Isabelle proof assistant, implementing a resolution prover and formalizing completeness of SPA's kernel. Better integration with Isabelle could help with inputting first-order formulas and giving better error-reporting when students make mistakes in formal proofs in SPA.

We have here considered proofs of Pelletier's problems 34 and 43 in SPA --- obviously only for advanced students --- but we are planning a new course on automated reasoning where SPA as well as 
a simple automatic prover \cite{VilladsenEtAl:PAAR2018} will play a central role. The key difficulty is to develop the necessary teaching materials. The tools and formalizations in the Isabelle proof assistant are now available.

\section*{Acknowledgements}

We thank Alexander Birch Jensen for collaboration on the first formalization and we thank Martin Elsman, Lars Hupel, John Bruntse Larsen and Makarius Wenzel for fruitful discussions.
We also thank the anonymous reviewers for their comments.

We are grateful to John Harrison for encouragement -- John once remarked to us that ``it is rather reassuring that I don't have to worry any longer about bugs in that part of the code'' given our formalization and code generation in Isabelle.

\bibliographystyle{eptcs}
\bibliography{references}

\newpage

\section*{Appendix: Complete Proof of Pelletier's Problem 34}

\begin{center}
\includegraphics[trim=0mm 30mm 0mm 5mm,clip,width=\textwidth,height=\textheight,keepaspectratio,page=1]{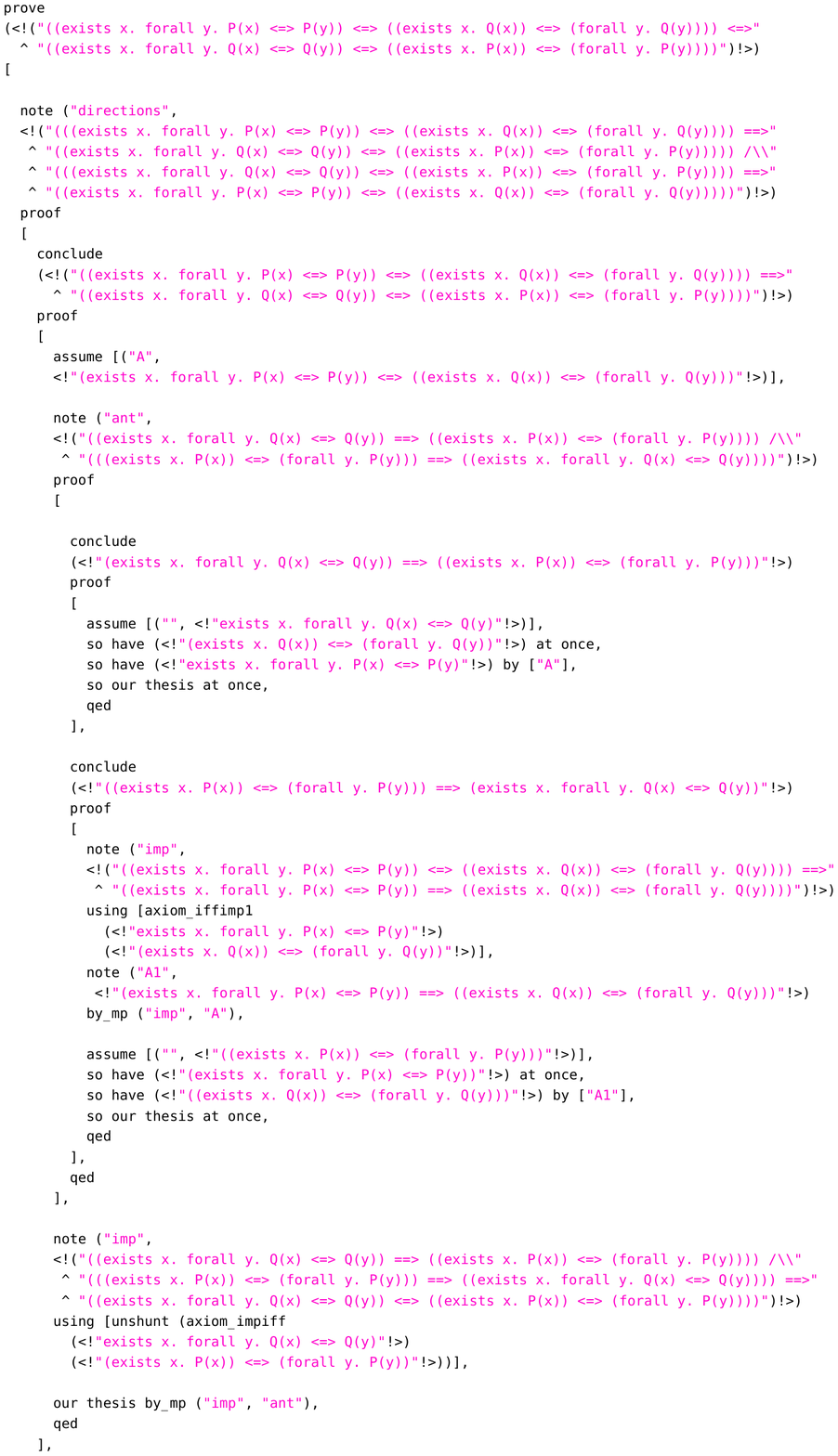}
\end{center}

%\vfill

%\noindent Continues.

\begin{center}
\includegraphics[trim=0mm 10mm 0mm 5mm,clip,width=\textwidth,height=\textheight,keepaspectratio,page=2]{SPA34.pdf}
\end{center}

\end{document}